\documentclass[aps,prl,twocolumn,superscriptaddress,longbibliography]{revtex4-2}
\usepackage{graphicx}
\usepackage{amssymb,amsmath}
\usepackage{bm}
\usepackage{dcolumn}
\usepackage{float}
\usepackage[OT1]{fontenc} 
\usepackage{url}
\usepackage{mathrsfs}
\usepackage{slashed,comment}
\usepackage{color}
\usepackage{verbatim}
\usepackage{enumitem}
\usepackage{soul,physics}
\usepackage[driverfallback=dvipdfm]{hyperref}
\hypersetup{pdfpagemode=FullScreen,colorlinks=true,breaklinks,urlcolor=blue,linkcolor=blue,citecolor=blue}

\usepackage{subfigure}
\usepackage{amssymb}
\usepackage{bm}
\usepackage{graphicx}
\usepackage{amsmath,amssymb}

\usepackage{pdfpages} 
\usepackage{pgffor} 

\makeatletter
\AtBeginDocument{\let\LS@rot\@undefined}
\makeatother

\def\supplementfilename{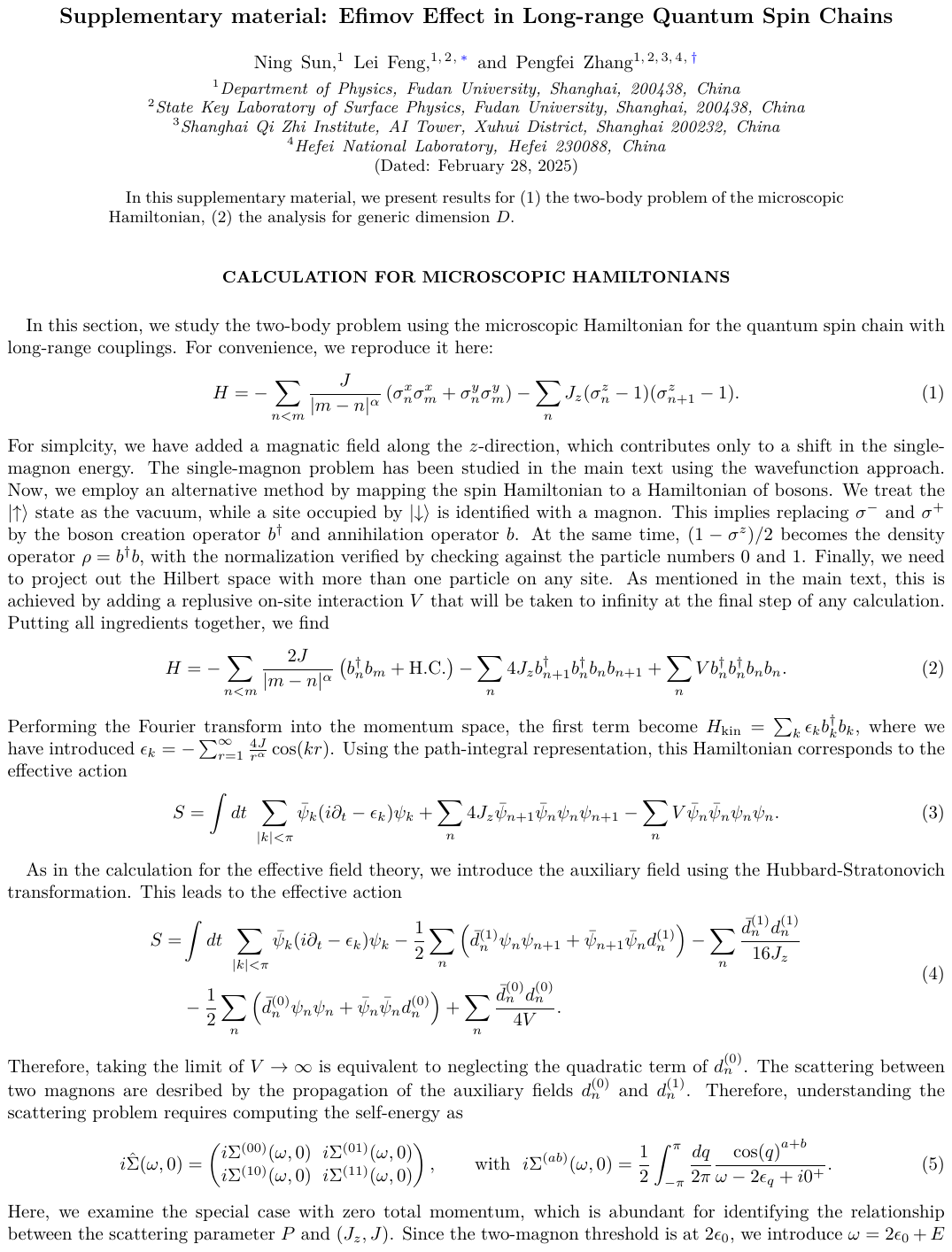}

\pdfximage{\supplementfilename}
\def\numbersupplementpages{\the\pdflastximagepages}

\newif\ifarXiv
\arXivtrue 

\begin{document}
 
  \title{Efimov Effect in Long-range Quantum Spin Chains }

  \author{Ning Sun}
  \affiliation{Department of Physics, Fudan University, Shanghai, 200438, China}

  \author{Lei Feng}
  \thanks{leifeng@fudan.edu.cn}
  \affiliation{Department of Physics, Fudan University, Shanghai, 200438, China}
  \affiliation{State Key Laboratory of Surface Physics, Fudan University, Shanghai, 200438, China}

  \author{Pengfei Zhang}
  \thanks{PengfeiZhang.physics@gmail.com}
  \affiliation{Department of Physics, Fudan University, Shanghai, 200438, China}
  \affiliation{State Key Laboratory of Surface Physics, Fudan University, Shanghai, 200438, China}
  \affiliation{Shanghai Qi Zhi Institute, AI Tower, Xuhui District, Shanghai 200232, China}
  \affiliation{Hefei National Laboratory, Hefei 230088, China}

  \date{\today}

  \begin{abstract}
  When two non-relativistic particles interact resonantly in three dimensions, an infinite tower of three-body bound states emerges, exhibiting a discrete scale invariance. This universal phenomenon, known as the Efimov effect, has garnered extensive attention across various fields, including atomic, nuclear, condensed matter, and particle physics. In this letter, we demonstrate that the Efimov effect also manifests in long-range quantum spin chains. The long-range coupling modifies the low-energy dispersion of magnons, enabling the emergence of continuous scale invariance for two-magnon states at resonance. This invariance is subsequently broken to discrete scale invariance upon imposing short-range boundary conditions for the three-magnon problem, leading to the celebrated Efimov bound states. Using effective field theory, we theoretically determine how the ratio of two successive binding energies depends on the interaction range, which agrees with the numerical solution of the bound-state problem. We further discuss generalizations to arbitrary spatial dimensions, where the traditional Efimov effect serves as a special case. Our results reveal universal physics in dilute quantum gases of magnons that can be experimentally tested in trapped-ion systems.
  \end{abstract}
    
  \maketitle

  \emph{ \color{blue}Introduction.--} Universality states that systems with distinct microscopic details can exhibit the same macroscopic physics. A famous example is the low-energy scattering of non-relativistic particles in three dimensions. In this case, the two-body physics is governed by a universal parameter, known as the scattering length \cite{landau2013quantum}. More intriguingly, when the system is tuned to two-body resonance, an infinite tower of three-body bound states can emerge, obeying discrete scale invariance $E_{n}/E_{n+1}=\exp\left({2\pi}/{s_0}\right)$ with $s_0\approx 1.00624$ for identical bosons. This is known as the Efimov effect, first discovered by Vitaly Efimov in 1970 \cite{EFIMOV1970563}. The physical origin of the Efimov effect is the emergence of continuous scaling symmetry in the two-body sector, which is subsequently broken to discrete scale invariance in three-body calculations. Due to its universality, the Efimov effect has attracted attention from various fields, including atomic, nuclear, condensed matter, and particle physics \cite{Hammer:2005bp,Nielsen:2001hbm,Ferlaino2010FortyYO,Braaten:2006vd,Hammer:2010kp,Nishida:2012hf,Naidon:2016dpf,sciadvaau5096,Kievsky:2021ghz,ZHANG2019289}. In particular, experimental observations of Efimov bound states have been reported in cold atom systems \cite{2006Natur.440..315K,PhysRevLett.112.190401,PhysRevLett.112.250404,PhysRevLett.113.240402} as well as in the helium trimer \cite{doi:10.1126/science.aaa5601}. Nevertheless, Efimov physics is largely restricted to three dimensions, while existing extensions to lower dimensions require considering mixed dimensions or multi-body interactions \cite{Nishida:2011ew,PhysRevLett.101.170401,PhysRevA.79.060701}.

  \begin{figure}[t]
    \centering
    \includegraphics[width=0.95\linewidth]{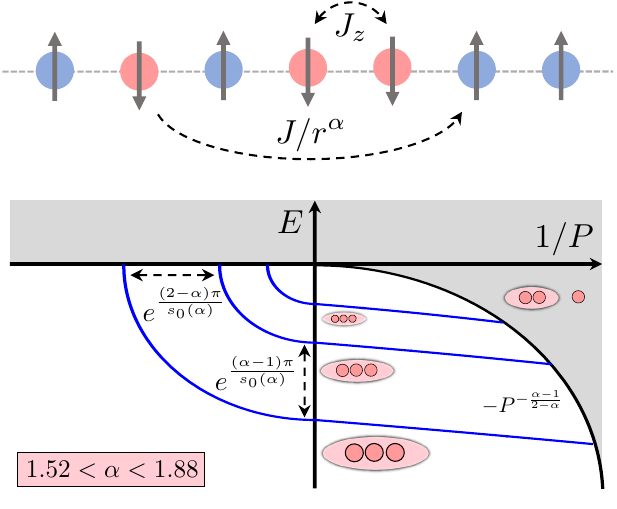}
    \caption{We present a schematic of our main results. We consider one-dimensional spin-1/2 quantum spin chains with long-range couplings in the $x/y-$direction, with spin-rotation symmetry along the $z$-direction. Here, we choose $\ket{\uparrow}$ as the vacuum and $\ket{\downarrow}$ represents magnons. When the decay exponent $\alpha\in(1.52,1.88)$, an infinite number of universal three-magnon bound states emerge, exhibiting discrete scale invariance with a scale factor that depends on $\alpha$. This phenomenon is known as the Efimov effect.  }
    \label{fig:schemticas}
  \end{figure}

  On the other hand, recent developments in quantum science and technology have significantly enlarged the classes of Hamiltonians realized in experiments. A prominent example is quantum spin chains with long-range couplings, which are relevant in trapped-ion systems \cite{wu2023research,2019ApPRv...6b1314B,Foss-Feig:2024blk,2021NatRM...6..892B,10164086,RevModPhys.93.025001}, Rydberg arrays \cite{Evered:2023aa,Ma:2023aa,Bluvstein:2024aa,Bekenstein:2020aa,Bluvstein:2021aa,Ebadi:2022aa,Bluvstein:2022aa,Lis:2023aa,Manetsch:2024aa,Tao:2024aa,Cao:2024aa}, and solid-state NMR systems \cite{JONES2001325,RevModPhys.76.1037,Lu2015NMRQI,Cory2000NMRBQ,Laflamme2002IntroductionTN}. In particular, in trapped-ion systems, the power of the coupling strength $J_{mn}\sim J/|m-n|^\alpha$ can be tuned for $\alpha\in(1,3)$ in state-of-the-art experiments through the precise control of spin-phonon interactions. This enables the observation of various interesting many-body phenomena, including symmetry breaking \cite{Kim:2010hll,Monroe:2011vee,Schneider2012ExperimentalQS}, topological phases \cite{PhysRevLett.107.150501,Dumitrescu:2021uin}, and dynamical behaviors \cite{Morong:2021qyr,Zhang:2017kde,Joshi:2021fno}. In this letter, we aim to emphasize that this system also serves as a novel platform for realizing universal few-body physics beyond the traditional paradigm. We focus on spin-1/2 systems with spin rotation symmetry along the $z$-direction, which ensures magnon number conservation. Unlike non-relativistic particles with quadratic dispersion, long-range coupling leads to a generic dispersion $\epsilon_k=u |k|^z$ with $z\in(0,2]$ for magnons \cite{LEPORI201635,PhysRevB.94.125121,PhysRevA.93.043605}, thereby significantly enriching the physics of low-energy scattering. In particular, we show that for $z\in (1/2,1)$, which corresponds to $\alpha \in(3/2,2)$, the two-body resonance of magnons leads to continuous scale invariance, similar to that of three-dimensional non-relativistic particles. We further demonstrate the existence of Efimov bound states in the three-magnon sector for $\alpha\in (1.52,1.88)$, with a tunable scale factor that depends on $\alpha$, as sketched in FIG. \ref{fig:schemticas}. Additionally, we generalize our results to long-range spin models in arbitrary dimension $D$, and discuss the implications for many-magnon states in the dilute limit.

  \emph{ \color{blue}Two-magnon Physics.--} 
  We study quantum spin chains with long-range interactions. Although our conclusions are valid for a broad class of models due to universality, we focus on a specific Hamiltonian for concreteness:
  \begin{equation}\label{eqn:microscopic}
  H=-\sum_{n<m}\frac{J}{|m-n|^\alpha}\left(\sigma_n^x\sigma_{m}^x+\sigma_n^y\sigma_{m}^y\right)-\sum_n{J_{z}}\sigma_n^z\sigma_{n+1}^z.
  \end{equation}
  Here, $\sigma_n^a$ with $a \in \{x,y,z\}$ represents the Pauli matrix on the $n$-th site. The Hamiltonian exhibits spin rotation symmetry along the $z$-direction, and the eigenstates are labeled by the eigenvalue of the total magnetization $\sum_n \sigma^z_n$. To draw an analogy with traditional scattering problems, we treat the fully polarized state $|0\rangle\equiv\otimes_n\ket{\uparrow}_n$ as a vacuum, and a down spin as a particle (magnon). The single-magnon state with momentum $k$ is given by $|k\rangle \sim \sum_n e^{ikn} \sigma^-_n |0\rangle$, with an excitation energy of
  \begin{equation}
  \epsilon_k=-\sum_{r=1}^\infty\frac{4J}{r^\alpha} \cos(kr)+4 J_{z}.
  \end{equation}
  Here, we have subtracted the energy of the vacuum state, $E_0=-\sum_{n,r>0}{J_{z,r}}$. The convergence of the summation for a generic $k$ requires $\alpha>1$. Focusing on the low-energy limit, an expansion at small $k$ gives \cite{LEPORI201635,PhysRevB.94.125121,PhysRevA.93.043605} 
  \begin{equation}\label{eq:dispersion}
  \epsilon_k-\epsilon_0\approx\begin{cases}-4 J\Gamma(1-\alpha)\sin \frac{\alpha \pi}{2} |k|^{\alpha-1} &  \alpha\in (1,3),\\
     2J{\zeta(\alpha-2)}k^2  &  \alpha \in [3,\infty).
    \end{cases}.
  \end{equation}
  Since the system conserves the magnon number, the constant term $\epsilon_0$ can always be subtracted by applying a homogeneous magnetic field, and will be dropped in the subsequent discussions. For conciseness, we denote the dispersion of magnons as $\epsilon_k=u |k|^z$, with a tunable dynamical exponent $z=\text{min}\{\alpha-1,2\}$ and $u>0$. 

  To proceed, we examine the low-energy properties of two-magnon states. The first term in \eqref{eqn:microscopic} describes the hopping of magnons, subject to the constraint that two magnons cannot occupy the same site, as $(\sigma_n^-)^2=0$. This is equivalent to considering ordinary particles with on-site repulsion $V$, and taking $V\rightarrow \infty$. For $J_z>0$, the second term in \eqref{eqn:microscopic} corresponds to an attractive potential when two magnons are in nearest neighbor sites. Therefore, we generally expect tunning $J_z$ will lead to a two-body resonance, where a new two-body bound state emerges. A detailed justification is provided in the supplementary material \cite{SM}. Following the general philosophy of low-energy effective field theory, the universal properties near this resonance can be captured by the Lagrangian \cite{Hammer:2005bp,zhai2021ultracold}
  \begin{equation}\label{eq:fieldtheory}
  L=\sum_k \bar{\psi}_k(i\partial_t-\epsilon_k )\psi_k-\frac{1}{2}\int dx~\Big(\bar{\psi}\bar{\psi}d+\bar{d}{\psi}{\psi}-\frac{\bar{d}d}{g}\Big).
  \end{equation}
  Here, $\psi$ denotes the magnon field. We have applied the Hubbard–Stratonovich transformation to introduce the dimer field $d$, which mediates the contact interaction between magnons \cite{Bedaque:1998kg, Bedaque:1998km}. The validity of Eq. \eqref{eq:fieldtheory} only requires the interaction between magnons to decays sufficiently rapidly, ensuring the validity of the short-range approximation \cite{landau2013quantum}. For instance, it can work even when the coupling in the $z$-direction is also long-range, analogous to the Van der Waals potential in cold atoms.
  
  \begin{figure}[t]
    \centering
    \includegraphics[width=0.92\linewidth]{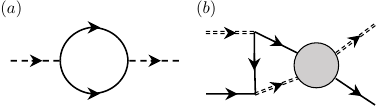}
    \caption{ (a) The self-energy diagram of the dimer field $d$. Solid and dashed lines represent the propagators of $\psi$ and $d$, respectively. (b) The scattering process between a dimer $d$ and a magnon $\psi$. The doubled dashed lines represent the renormalized propagator of the dimer field $d$. }
    \label{fig:diagrams}
  \end{figure}
  
  We consider the scattering between two incoming magnons with total energy $E$ and total momentum $k$. The scattering $T$-matrix $T(E,k)$ matches the full dimer propagator with the self-energy diagram shown in FIG. \ref{fig:diagrams} (a). This leads to
  \begin{equation}\label{eqn:two-body}
  T(E,k)^{-1}=\frac{1}{2g}-\frac{1}{2}\int^\Lambda_{-\Lambda} \frac{dq}{2\pi}\frac{1}{E_+-\epsilon_{\frac k2+q}-\epsilon_{\frac k2-q}}.
  \end{equation}
  Here, we introduced $E_+=E+i0^+$ and regularized the integral using a high-momentum cutoff $\Lambda$. Important insights into the two-body scattering can be obtained by analyzing the the integral at $k=0$: The integrand scales as $q^{-z}+Eq^{-2z}+...$ in the limit of large $q$. For $z>1$, the integral converges. This situation is similar to non-relativistic particles in one dimension, where the system is weakly interacting ($T(E,0)\rightarrow 0$ as $E\rightarrow 0$) near the resonance. For $z<1/2$, the integral exhibits multiple divergent terms, and its renormalization requires additional kinetic terms for the dimer fields. However, this would introduce additional dimensionful parameters, thereby breaking scale invariance at resonance. Therefore, a non-trivial two-magnon resonance with continuous scale invariance can only exist for $z\in(1/2,1)$, which corresponds to $\alpha \in (3/2,2)$. We find
  \begin{equation}
   T(E,0)^{-1}={P}^{-1}-C_0u^{-\frac{1}{z}}(-E_+)^{\frac{1-z}{z}}.
  \end{equation}
  where $C_0=-{2^{-\frac{1}{z}-1} /{z\sin \left(\frac{\pi }{z}\right)}}$ is a positive constant and the scattering parameter $P$ is introduced by the renormalization relation
  \begin{equation}
  \frac{1}{P}= \frac{1}{2g}+\frac{\Lambda^{1-z}}{4\pi u(1-z)}.
  \end{equation} 

  It is straightforward to verify that this renormalization relation holds at finite $k$, although closed-form expressions for $T(E,k)$ are no longer available. Similar to the scattering length in three dimensions, the scattering parameter $P$ governs the low-energy physics in the two-body sector. When $P>0$, the $T$-matrix contains a pole at $E=-u^{\frac{1}{1-z}}(PC_0)^{-\frac{z}{1-z}}$. This is the energy of the two-body bound state, which decreases monotonically as $1/P$ is tunned from $0^+$ to $+\infty$, as illustrated in FIG. \ref{fig:schemticas}. The two-body resonant occurs at $P=\infty$, where magnons interact strongly with a scale-invariant $T$-matrix. This opens up the possibility for the emergence of Efimov bound states in the three-magnon sector. In the supplementary material \cite{SM}, we further provide the direct relation between $P$ and parameters in the microscopic Hamiltonian \eqref{eqn:microscopic}. 
  
  \begin{figure}[t]
    \centering
    \includegraphics[width=0.85\linewidth]{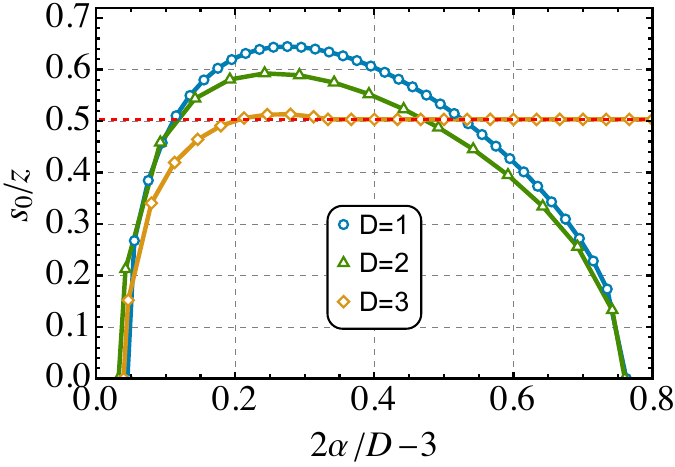}
    \caption{ The theoretical prediction of the scale factor for Efimov states $E_n=E_0 \exp({-\frac{z\pi}{s_0(\alpha)}}n) $ in different dimensions as a function of $\alpha$. Here, we have dynamical exponent $z=\text{min}\{\alpha-D,2\}$. The Efimov states exist for $\alpha \in (1.52,1.88)$ in one dimension, for $\alpha \in (3.03,3.76)$ in two dimensions and for $\alpha>4.56$ in three dimensions. The red dashed line represents the result for non-relativistic particles in three dimensions, which matches the prediction for three dimensional long-range spin models with $\alpha\geq 5$. }
    \label{fig:res}
  \end{figure}

  \emph{ \color{blue}Efimov Effect.--} Now, we are ready to study the three-magnon problem at $P=\infty$ by considering one incoming dimer and one incoming magnon. The relevant diagram is shown in FIG. \ref{fig:diagrams} (b), where the doubled dashed lines represent the renormalized propagator of dimers. Focusing on the bound-state problem, we set the total energy $E<0$ and momentum $k=0$. The bound-state wavefunction $\varphi(E,k)$ satisfies the celebrated Skorniakov-Ter-Martirosian (STM) equation \cite{Skorniakov:1957kgi}
  \begin{equation}\label{eqn:STM}
  \varphi(E,k)=\int_{-\Lambda}^\Lambda \frac{dq}{2\pi}\frac{T(E-\epsilon_q,-q)}{E-\epsilon_k-\epsilon_q-\epsilon_{k-q}}\varphi(E,q).
  \end{equation}
  The values of $E$ for which this homogeneous integral equation has solutions correspond to the energies $-E_n$ of the three-magnon bound states. We first investigate the zero-energy wave function $\varphi(k)\equiv \varphi(0,k)$ with $\Lambda \rightarrow \infty$, which captures the universal behavior of $\phi(E,k)$ at $\Lambda \gg k\gg (-E)^{1/z}$. It satisfies the integral equation 
  \begin{equation}\label{eq:STMzeroE}
  \varphi(k)=\int_{-\infty}^\infty \frac{dq}{2\pi}\frac{T(-\epsilon_q,-q)}{-\epsilon_k-\epsilon_q-\epsilon_{k-q}}\varphi(q).
  \end{equation}
  At the two-magnon resonance $P=\infty$, this equation is invariant under the scale transformation $(k,q)\rightarrow \lambda (k,q)$, since Eq. \eqref{eqn:two-body} predicts $T(-\epsilon_q,-q)=-u/|q|^{1-z}B(z)$ with a numerical factor 
  \begin{equation}\label{eq:integral1}
   B(z)\equiv\int_0^\infty \frac{dx}{2\pi} \left(\frac{1}{2x^z}-\frac{1}{1+(\frac{1}{2}+x)^z+|\frac{1}{2}-x|^z}\right).
  \end{equation}
  Therefore, its solution generally takes the form of $\varphi(q)\sim |q|^{s_1\pm is_0}$ \footnote{Here, we focus on the even-wave solution between magnon and dimer. }. The parameter $s_1$ and $s_0$ are determined by substituting the ansatz into Eq. \eqref{eq:STMzeroE}. This yields $s_1=-\frac{z}{2}$, while $s_0(\alpha)$ satisfies the integral equation 
  \begin{equation}\label{eq:integral2}
  \int_{-\infty}^\infty \frac{dx}{2\pi}\frac{|x|^{\frac{z}{2}-1+is_0}}{1+|x|^z+|1-x|^z}=B(z).
  \end{equation}

\begin{table}[t]
  \centering 
  \begin{tabular}{|c|c|c|c|c|c|}
    \hline
    \hline
    &\multicolumn{2}{c|}{$z$=0.65} &\multicolumn{2}{c|}{$z$=0.75}\\ 
    \hline
    $n$ & $\phi_n$  & $\phi_n-\phi_{n-1}$& $\phi_n$  & $\phi_n-\phi_{n-1}$ \\ 
    \hline
    1  & 2.203 &   & 2.713 &    \\ 
    2 & 3.761 &1.558 &  4.606 &1.893\\ 
    3  & 5.318 &1.557 & 6.497 &1.891  \\ 
    4  & 6.875&1.557 &  8.388 &1.891  \\ 
   \hline
    $z/s_0$ & & 1.556& & 1.892 \\
   \hline   
    \end{tabular}
  
  \caption{The numerical results for the three-body binding energies $E_n$ by solving Eq. \eqref{eqn:STM} with $E_n=u\Lambda^{z}\exp(-\pi\phi_n)$. The differences $\phi_{n}-\phi_{n-1}$ match the theoretical prediction of $z/s_0$ with good accuracy.  }\label{TAB:numeric}
\end{table}

  Unfortunately, we are unable to obtain an analytical solution for general $z$. Instead, we solve Eq. \eqref{eq:integral1} and Eq. \eqref{eq:integral2} numerically. The result is shown in FIG. \ref{fig:res} using blue circles. The solution of $s_0$ exists only for $z \in (0.52, 0.88)$, which corresponds to $\alpha \in (1.52,1.88)$. In this parameter regime, a superposition of $|q|^{s_1 \pm is_0}$ leads to $\varphi (q)\sim |q|^{s_1}\cos(s_0\ln|q|/\Lambda_*)$. Here, $\Lambda_*$ is the three-body parameter, which is now sensitive to the cutoff $\Lambda$ in Eq. \eqref{eqn:STM}. This log-periodic wavefunction implies the breaking of continuous scale invariance down to discrete scale invariance \cite{Hammer:2005bp}: Under a scale transformation $q\rightarrow \lambda q$, the wavefunction is invariant (up to a global phase) only if $\lambda=\exp(\frac{\pi}{s_0(\alpha)}n)$, where $n$ is an integer. This implies all three-magnon bound states organize into a geometric series with $E=-E_n=-E_0 \exp({-\frac{z\pi}{s_0(\alpha)}}n) $ and $z=\alpha-1$, which is the infinite tower of Efimov states. Generalizing this argument to include a small but finite $P$, we obtain the schematic shown in FIG. \eqref{fig:schemticas}. Interestingly, long-range spin chains can achieve a energy ratio $e^{z\pi/s_0(\alpha)}\approx130$ near $\alpha=1.63 $, which is much smaller than that for identical bosons in three dimensions $e^{2\pi/s_0}\approx515$. To validate our theoretical prediction, we have calculated the binding energies by solving the STM equation \eqref{eqn:STM} numerically. The results for $z=0.65$ and $z=0.75$ are presented in TABLE \ref{TAB:numeric}.

  The presence of Efimov states indicates that an additional renormalization relation is necessary to eliminate the cutoff dependence \cite{Bedaque:1998kg, Bedaque:1998km}. Here, we add a contact interaction between magnon and dimer $\Delta L= -h\int dx~\bar{\psi}\bar{d}d\psi$, where $h(\Lambda)=-H(\Lambda)/u\Lambda^z$ by dimensional counting. Adding the contribution from this new vertex to \eqref{eqn:STM} and focusing on the integral near $q\approx \Lambda$, the R.H.S. becomes
  \begin{equation}
  \propto \int^\Lambda dq  \left[\frac{1}{2q^{1+\frac z2}}+\frac{H(\Lambda)}{\Lambda^zq^{1-\frac z2}}\right]\cos(s_0\ln q/\Lambda_*).
  \end{equation}
  Requiring the cancellation of the cutoff dependence, we obtain the three-body renormalization relation
  \begin{equation}
  H(\Lambda)=\frac{\cos(s_0\ln (\Lambda/\Lambda_*)+\arctan(2s_0/z))}{\cos(s_0\ln (\Lambda/\Lambda_*)-\arctan(2s_0/z))}.
  \end{equation}
  This renormalization relation ensures the bound state energy is cutoff independent and has broad implication for universal properties of the many-magnon states \cite{PhysRevLett.112.110402,PhysRevLett.106.153005,PhysRevA.86.053633}. 

  \emph{ \color{blue}Higher Dimensions.--} By generalizing our calculations to arbitrary spatial dimension $D$, we show that the Efimov effect also occurs for $D=2,3$. Here, we provide an outline of the analysis and refer to the supplementary material \cite{SM} for the detailed calculations. 

  Our microscopic model still contains with long-range XY coupling with an exponent $\alpha$ and a short-range coupling in the $z$-direction. Similar to Eq. \eqref{eq:dispersion}, the single-magnon state exhibits a tunable dynamical exponent $\epsilon_\mathbf{k}=u |\mathbf{k}|^z$, but with $z=\text{min}\{\alpha-D,2\}$ for $\alpha>D$. Next, we identify the regime where the self-energy of the dimer fields contains a single divergent term, which is renormalized to the only dimensionful scattering parameter. The difference compared to the one-dimensional case is that the integral in Eq. \eqref{eqn:two-body} is now $D$-dimensional. This selects $z\in (D/2,D)$ in general dimensions. For $D=2$, this leads to $z\in (1,2)$, which corresponds to $\alpha\in (3,4)$. For $D=3$, it requires $z\in(3/2,2]$, or equivalently, $\alpha>9/2$. In this case, the physics is exactly the same as non-relativistic particles for $\alpha>5$ \cite{Nishida:2012hf}. Finally, for $D\geq 4$, the conditions are not compatible, and we expect the absence of Efimov effect for arbitrary $\alpha$. 
  
  With this understanding, we further study three-magnon bound states using the STM equation in $D=2,3$. The predictions for the scaling factors are also presented in Fig. \ref{fig:res}, with green triangles ($D$=2) and yellow diamonds ($D=3$). The results for $D=2$ shows similar features to $D=1$, exhibiting the Efimov effect when $\alpha$ lies in a finite region $(3.03,3.76)$. Systems with $D=3$ display the Efimov effect when $\alpha>4.56$, consistent with the presence of the Efimov effects in identical bosons. Interestingly, when expressed in terms of $2\alpha/D-3$, the upper bounds for $D=1$ and $D=2$ are quite close to each other. However, numerical calculations still reveal a non-zero discrepancy. The absence of Efimov effect close to $\alpha=D/2$ is also consistent with the stability of resonant bose gas proposed in \cite{PhysRevA.93.043605}. 

  \emph{ \color{blue}Discussions.--} In this letter, we demonstrate that long-range spin models support an infinite tower of universal three-magnon bound states, with the binding energies organized in a geometric series across a finite range of system parameters $\alpha$. This phenomenon generalizes the celebrated Efimov effect, initially observed in non-relativistic particles in three dimensions. Our findings offer a new pathway for observing the Efimov effect in low dimensions and open new avenues for investigating universal few-body physics with a general dynamical exponent across various quantum platforms. In particular, our theoretical prediction aligns perfectly with recent advances in trapped-ion systems \cite{wu2023research, 2019ApPRv...6b1314B, Foss-Feig:2024blk, 2021NatRM...6..892B, 10164086, RevModPhys.93.025001}, where precise control of the parameter $\alpha$ is achievable.

  We conclude our work with a few remarks on future directions. Firstly, it is possible to further extend the parameter regime for observing the Efimov effect by considering scattering problems in mixed dimensions \cite{Nishida:2011ew,PhysRevLett.101.170401,PhysRevA.79.060701}, which corresponds to an impurity problem in quantum spin models. Secondly, novel universal bound states beyond the Efimov paradigm exist, such as the super-Efimov effect \cite{PhysRevLett.110.235301,PhysRevA.90.063631,2014arXiv1405.1787G,PhysRevA.92.020504,PhysRevA.95.033611} and the semi-super-Efimov effect \cite{PhysRevLett.118.230601,PhysRevA.96.030702}. Whether these states also emerge, possibly for a higher partial wave, in long-range spin chains presents an intriguing problem. Finally, universal few-magnon states have direct implications when the magnon density $n_m$ is small but finite, similar to dilute quantum gases. In this regime, the physics remains dominated by low-energy scattering theory, and a set of powerful relation \cite{TAN20082971,PhysRevLett.100.205301,PhysRevLett.99.190407,PhysRevLett.99.170404,PhysRevLett.116.045301,2009EPJB...68..401W,PhysRevA.79.053640,PhysRevA.79.023601}, known as Tan's contact relations holds. In our setup, we can introduce two- and three-body contacts as $C_2=-\partial \langle H\rangle /\partial(P^{-1})$ and $C_3=\Lambda_*\partial \langle H\rangle /\partial(\Lambda_*)$. For example, the renormalization relation indicates the momentum distribution of magnons $n(k)=\langle \bar{\psi}_k\psi_k\rangle$ for $n_m\ll k\ll 1$ can be expanded as
  \begin{equation}
  n(k)\sim \frac{1}{4u^2 k^{2z}}C_2+\frac{F(\ln (k/\Lambda_*))}{uk^{1+z}}C_3+...
  \end{equation}
  Here, $F(x)$ is a universal function with periodicity $\pi/s_0$. Its derivation, together with the results of other universal relations, will be presented elsewhere.

\vspace{5pt}
\textit{Acknowledgement.} 
We thank Zhenhua Yu for helpful discussions. This project is supported by the Innovation Program for Quantum Science and Technology ZD0220240101 (PZ) and 2023ZD0300900 (LF), the Shanghai Rising-Star Program under grant number 24QA2700300 (PZ), the NSFC under grant 12374477 (PZ), and Shanghai Municipal Science and Technology Major Project grant 24DP2600100 (NS and LF).

\bibliography{Efimov_Long-range.bbl}

\ifarXiv


\section*{Supplementary material}

# Supplementary material: Efimov Effect in Long-range Quantum Spin Chains

Ning Sun,[1] Lei Feng,[1, 2, *] and Pengfei Zhang[1, 2, 3, 4, †]

[1]*Department of Physics, Fudan University, Shanghai, 200438, China*
[2]*State Key Laboratory of Surface Physics, Fudan University, Shanghai, 200438, China*
[3]*Shanghai Qi Zhi Institute, AI Tower, Xuhui District, Shanghai 200232, China*
[4]*Hefei National Laboratory, Hefei 230088, China*
(Dated: February 28, 2025)

In this supplementary material, we present results for (1) the two-body problem of the microscopic Hamiltonian, (2) the analysis for generic dimension $D$.

## CALCULATION FOR MICROSCOPIC HAMILTONIANS

In this section, we study the two-body problem using the microscopic Hamiltonian for the quantum spin chain with long-range couplings. For convenience, we reproduce it here:

$$H = -\sum_{n<m} \frac{J}{|m-n|^{\alpha}} \left(\sigma_n^x \sigma_m^x + \sigma_n^y \sigma_m^y\right) - \sum_n J_z(\sigma_n^z - 1)(\sigma_{n+1}^z - 1). \tag{1}$$

For simplcity, we have added a magnatic field along the $z$-direction, which contributes only to a shift in the single-magnon energy. The single-magnon problem has been studied in the main text using the wavefunction approach. Now, we employ an alternative method by mapping the spin Hamiltonian to a Hamiltonian of bosons. We treat the $|\uparrow\rangle$ state as the vacuum, while a site occupied by $|\downarrow\rangle$ is identified with a magnon. This implies replacing $\sigma^-$ and $\sigma^+$ by the boson creation operator $b^{\dagger}$ and annihilation operator $b$. At the same time, $(1 - \sigma^z)/2$ becomes the density operator $\rho = b^{\dagger}b$, with the normalization verified by checking against the particle numbers 0 and 1. Finally, we need to project out the Hilbert space with more than one particle on any site. As mentioned in the main text, this is achieved by adding a replusive on-site interaction $V$ that will be taken to infinity at the final step of any calculation. Putting all ingredients together, we find

$$H = -\sum_{n<m} \frac{2J}{|m-n|^{\alpha}} \left(b_n^{\dagger} b_m + \text{H.C.}\right) - \sum_n 4J_z b_{n+1}^{\dagger} b_n^{\dagger} b_n b_{n+1} + \sum_n V b_n^{\dagger} b_n^{\dagger} b_n b_n. \tag{2}$$

Performing the Fourier transform into the momentum space, the first term become $H_{\text{kin}} = \sum_k \epsilon_k b_k^{\dagger} b_k$, where we have introduced $\epsilon_k = -\sum_{r=1}^{\infty} \frac{4J}{r^{\alpha}} \cos(kr)$. Using the path-integral representation, this Hamiltonian corresponds to the effective action

$$S = \int dt \sum_{|k|<\pi} \bar{\psi}_k(i\partial_t - \epsilon_k)\psi_k + \sum_n 4J_z \bar{\psi}_{n+1} \bar{\psi}_n \psi_n \psi_{n+1} - \sum_n V \bar{\psi}_n \bar{\psi}_n \psi_n \psi_n. \tag{3}$$

As in the calculation for the effective field theory, we introduce the auxiliary field using the Hubbard-Stratonovich transformation. This leads to the effective action

$$S = \int dt \sum_{|k|<\pi} \bar{\psi}_k(i\partial_t - \epsilon_k)\psi_k - \frac{1}{2} \sum_n \left(\bar{d}_n^{(1)} \psi_n \psi_{n+1} + \bar{\psi}_{n+1} \bar{\psi}_n d_n^{(1)}\right) - \sum_n \frac{\bar{d}_n^{(1)} d_n^{(1)}}{16 J_z}$$
$$- \frac{1}{2} \sum_n \left(\bar{d}_n^{(0)} \psi_n \psi_n + \bar{\psi}_n \bar{\psi}_n d_n^{(0)}\right) + \sum_n \frac{\bar{d}_n^{(0)} d_n^{(0)}}{4V}. \tag{4}$$

Therefore, taking the limit of $V \to \infty$ is equivalent to neglecting the quadratic term of $d_n^{(0)}$. The scattering between two magnons are desribed by the propagation of the auxiliary fields $d_n^{(0)}$ and $d_n^{(1)}$. Therefore, understanding the scattering problem requires computing the self-energy as

$$i\hat{\Sigma}(\omega, 0) = \begin{pmatrix} i\Sigma^{(00)}(\omega, 0) & i\Sigma^{(01)}(\omega, 0) \\ i\Sigma^{(10)}(\omega, 0) & i\Sigma^{(11)}(\omega, 0) \end{pmatrix}, \qquad \text{with} \quad i\Sigma^{(ab)}(\omega, 0) = \frac{1}{2} \int_{-\pi}^{\pi} \frac{dq}{2\pi} \frac{\cos(q)^{a+b}}{\omega - 2\epsilon_q + i0^+}. \tag{5}$$

Here, we examine the special case with zero total momentum, which is abundant for identifying the relationship between the scattering parameter $P$ and $(J_z, J)$. Since the two-magnon threshold is at $2\epsilon_0$, we introduce $\omega = 2\epsilon_0 + E$

and focus on the low-energy scattering at small $|E| \ll J$. The important observation is that, the leading-order behavior of a small but finite $E$ comes from the integration near $q \approx 0$, where the integral can be approximated as

$$i\Sigma^{(ab)}(\omega, 0) \approx i\Sigma^{(ab)}(0,0) + \frac{1}{2}\int_{-\infty}^{\infty}\frac{dq}{(2\pi)}\left[\frac{1}{E - 2u|q|^z + i0^+} + \frac{1}{2u|q|^z}\right] + o(E^{\frac{1-z}{z}}). \tag{6}$$

Consequently, the second term matches the result of the effective field theory presented in the main text and we have

$$i\Sigma^{(ab)}(\omega, 0) \approx i\Sigma^{(ab)}(0,0) + C_0 u^{-\frac{1}{z}}(-E_+)^{\frac{1-z}{z}} \tag{7}$$

with $C_0 = -2^{-\frac{1}{z}-1}/z \sin\left(\frac{\pi}{z}\right)$. This leads to the inverse propagator of the auxiliary field:

$$\left[-i\hat{G}(\omega, 0)\right]^{-1} \approx \begin{pmatrix} -i\Sigma^{(00)}(0,0) - C_0 u^{-\frac{1}{z}}(-E_+)^{\frac{1-z}{z}} & -i\Sigma^{(10)}(0,0) - C_0 u^{-\frac{1}{z}}(-E_+)^{\frac{1-z}{z}} \\ -i\Sigma^{(10)}(0,0) - C_0 u^{-\frac{1}{z}}(-E_+)^{\frac{1-z}{z}} & -\frac{1}{16J_z} - i\Sigma^{(11)}(0,0) - C_0 u^{-\frac{1}{z}}(-E_+)^{\frac{1-z}{z}} \end{pmatrix}. \tag{8}$$

Here, $-i\Sigma^{(ab)}(0,0) = C^{ab}/J$, with numerical factors $C^{ab} > 0$ determined by the numerical integration of

$$C^{ab} = \int_{-\pi}^{\pi}\frac{dq}{8\pi}\frac{\cos(q)^{a+b}}{\sum_{r=1}^{\infty}\frac{4}{r^\alpha}[1 - \cos(qr)]}. \tag{9}$$

For example, we have $(C^{00}, C^{10}, C^{11}) \approx (0.0617, 0.0354, 0.0459)$ for $z = 0.75$ or $\alpha = 1.75$.

The resonent point is determined by looking for the emergent of a bound state at $E = 0$, which corresponds to a pole at $\omega = 0$, detected by $\det\left[\left[-i\hat{G}(0,0)\right]^{-1}\right] = 0$. We find

$$\frac{J_z^*}{J} = \frac{C^{00}}{16(C^{00}C^{11} - (C^{01})^2)} > 0. \tag{10}$$

Here, the dinominator can be proven positive using the Cauchy-Schwarz inequality. Expanding near $\delta J_z = J_z - J_z^* \ll J$, the propagator is dominated by

$$\left[-i\hat{G}(\omega, 0)\right]_{ab}^{-1} \approx v_a v_b \left\{\frac{16\left[C^{00}C^{11} - (C^{01})^2\right]^2}{(C^{00})^2 + (C^{01})^2}\frac{\delta J_z}{J} - C_0 u^{-\frac{1}{z}}(-E_+)^{\frac{1-z}{z}}\left[1 - \frac{2C^{00}C^{01}}{(C^{00})^2 + (C^{01})^2}\right]\right\} \tag{11}$$

Here, $\boldsymbol{v} = \frac{(-C^{01}, C^{00})}{\sqrt{(C^{01})^2 + (C^{00})^2}}$ is the eigenvector at $J_z = J_z^*$. Comparing to Eq. (6) in the main text, we immediately identify the relation

$$\frac{1}{P} = \frac{16\left[C^{00}C^{11} - (C^{01})^2\right]^2}{(C^{00})^2 + (C^{01})^2 - 2C^{00}C^{01}}\frac{\delta J_z}{J}. \tag{12}$$

In summary, the microscopic model involves two auxiliary fields. However, only a specific combination of these fields (determined by the eigenvector $\boldsymbol{v}$) contributes to the low-energy physics, which is identified with the dimer field $d$ in the effective field theory, and whose propagator matches Eq. (6) in the main text. This procedure leads, for example, to the same STM equation as Eq. (8) in the main text. Therefore, our derivation generally implies the validity of the effective field theory for generic few-body problems, including the three-body calculation in which Efimov bound states may arise.

## ANALYSIS IN GENERAL DIMENSIONS

In this section, we present our results in general dimensions $D$. We begin with the microscopic Hamiltonian for a quantum spin model with long-range couplings, given by

$$H = -\frac{1}{2}\sum_{\boldsymbol{r}\neq\boldsymbol{r}'}\frac{J}{|\boldsymbol{r}-\boldsymbol{r}'|^\alpha}(\sigma_{\boldsymbol{r}}^x\sigma_{\boldsymbol{r}'}^x + \sigma_{\boldsymbol{r}}^y\sigma_{\boldsymbol{r}'}^y) - \sum_{\langle\boldsymbol{r},\boldsymbol{r}'\rangle}J_z\sigma_{\boldsymbol{r}}^z\sigma_{\boldsymbol{r}'}^z, \tag{13}$$

Here, the notation $\sum_{\langle \boldsymbol{r}, \boldsymbol{r}' \rangle}$ represents the summation over nearest-neighbor sites. We focus on the square lattice case with $\boldsymbol{r} = (r_x, r_y)$ and $r_x, r_y \in \mathbb{Z}$. The single magnon state $|\boldsymbol{k}\rangle \sim \sum_{\boldsymbol{r}} e^{i\boldsymbol{k}\cdot\boldsymbol{r}} \sigma_{\boldsymbol{r}}^- |0\rangle$ has an excitation energy of

$$\epsilon_k = -\sum_{\boldsymbol{r}\neq\boldsymbol{0}}^{\infty} \frac{4J}{r^\alpha} \cos(\boldsymbol{k}\cdot\boldsymbol{r}) + 4DJ_z. \tag{14}$$

A finite energy requires $\alpha > D$. Similar to the one-dimensional case, a small-$k$ expansion gives

$$\epsilon_k = \epsilon_0 + u|\boldsymbol{k}|^z, \qquad \text{with} \quad z = \min\{2, \alpha - D\}. \tag{15}$$

As discussed in the main text, the interactions between magnons are short-ranged. As a consequence, we formulate an effective field theory to describe the low-energy scattering between magnons. The effective action is given by

$$S = \int dt \, \sum_{\boldsymbol{k}} \bar{\psi}_{\boldsymbol{k}}(i\partial_t - \epsilon_k)\psi_{\boldsymbol{k}} - \frac{1}{2}\int d\boldsymbol{r} \, \left( \bar{\psi}\bar{\psi}d + \bar{d}\psi\psi - \frac{\bar{d}d}{g} \right). \tag{16}$$

The scattering $T$-matrix $T(E, \boldsymbol{k})$ between two incoming magnons with total energy $E$ and total momentum $k$ matches the full propagator of the dimer field. We have

$$T(E, \boldsymbol{k})^{-1} = \frac{1}{2g} - i\Sigma(E, \boldsymbol{k}) = \frac{1}{2g} - \frac{1}{2}\int_{q<\Lambda} \frac{d\boldsymbol{q}}{(2\pi)^D} \frac{1}{E + -\epsilon_{\frac{\boldsymbol{k}}{2}+\boldsymbol{q}} - \epsilon_{\frac{\boldsymbol{k}}{2}-\boldsymbol{q}}}. \tag{17}$$

Having a strongly interacting, scale-invariant two-body resonance requires a single divergent term in the integral. This condition is satisfied for $\alpha \in (D/2, D)$. Under this constraint, we find

$$i\Sigma(E, \boldsymbol{0}) = -\frac{S_D}{(2\pi)^D 4u(D-z)}\Lambda^{D-z} - \frac{S_D\pi(2u)^{-D/z}\csc(D\pi/z)}{(2\pi)^D 2z}(-E)^{\frac{D-z}{z}}. \tag{18}$$

Here, $S_D = 2\pi^{D/2}/\Gamma(D/2)$ is the area of a $D-1$ dimensional unit sphere. This leads to the renormalization relation

$$\frac{1}{2g} + \frac{S_D}{(2\pi)^D 4u(D-z)}\Lambda^{D-z} \equiv \frac{1}{P}. \tag{19}$$

At $P \to \infty$, the system exhibits scale invariance. For later convenience, we investigate the propagator with momentum $\boldsymbol{k}$ and energy $E = -\epsilon_k$. The result is

$$T(-\epsilon_k, \boldsymbol{k})^{-1} = \frac{1}{2}\int \frac{d\boldsymbol{q}}{(2\pi)^D}\left[ \frac{1}{\epsilon_k + \epsilon_{\frac{\boldsymbol{k}}{2}+\boldsymbol{q}} + \epsilon_{\frac{\boldsymbol{k}}{2}-\boldsymbol{q}}} - \frac{1}{2\epsilon_{\boldsymbol{q}}} \right] \equiv -\frac{|\boldsymbol{k}|^{D-z}}{u}B(z, D). \tag{20}$$

Here, it is straightforward to check that $B(z, 1) = B(z)$, as introduced in the main text. More explicitly, we have

$$B(z, D) = -\frac{1}{2}\int \frac{d\boldsymbol{x}}{(2\pi)^D}\left[ \frac{1}{1 + |\boldsymbol{x} + \hat{e}/2|^z + |\boldsymbol{x} - \hat{e}/2|^z} - \frac{1}{2|\boldsymbol{x}|^z} \right], \tag{21}$$

with an arbitrary unit vector $\hat{e}$.

Now, we consider the three-magnon problem. The STM equation for the bound-state problem is given by

$$\varphi(E, \boldsymbol{k}) = \int \frac{d\boldsymbol{q}}{(2\pi)^D} \frac{T(E - \epsilon_q, -\boldsymbol{q})}{E - \epsilon_k - \epsilon_q - \epsilon_{\boldsymbol{k}-\boldsymbol{q}}}\varphi(E, \boldsymbol{q}). \tag{22}$$

We focus on the bound state problem in the sector with zero total angular momentum, where the bound-state wavefunction $\varphi(E, \boldsymbol{k})$ only depends on $|\boldsymbol{k}|$. Similar to the discussions in the main text, we examine the zero-energy wavefunction, which satisfies

$$\varphi(\boldsymbol{k}) = \int \frac{d\boldsymbol{q}}{(2\pi)^D} \frac{T(-\epsilon_q, -\boldsymbol{q})}{-\epsilon_k - \epsilon_q - \epsilon_{\boldsymbol{k}-\boldsymbol{q}}}\varphi(\boldsymbol{q}). \tag{23}$$

The solution is given by $\varphi(\boldsymbol{q}) \sim |\boldsymbol{q}|^{-\frac{z}{2}\pm is_0}$, with $s_0$ determined by the integral

$$\int \frac{d\boldsymbol{x}}{(2\pi)^D} \frac{x^{\frac{z}{2}-D+is_0}}{1 + x^z + |\boldsymbol{x} - \hat{e}|^z} = B(z, D). \tag{24}$$

Solving this equation numerically, we obtain Figure 3 presented in the main text.

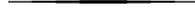

* leifeng@fudan.edu.cn
† PengfeiZhang.physics@gmail.com


\fi

\end{document}